\documentclass{PoS}
\usepackage{graphicx,subfigure}
\usepackage{amsmath,amssymb,amsfonts}
\usepackage{bm}
\usepackage{color}
\usepackage{comment}
\usepackage{slashed}

\def \be {\begin{equation} }
\def \ee {\end{equation}}

\def \bem {\begin{multline}}
\def \eem {\end{multline}}

\def \bes {\begin{subequations} }
\def \ees {\end{subequations}}

\newcommand{\Eq}[1]{Eq.~(\ref{#1})}

\title{World-line formulation of chiral kinetic theory in topological background gauge fields}

\ShortTitle{World-line approach to chiral kinetic theory}

\author{Niklas Mueller\\
        Physics Department, Brookhaven National Laboratory, Bldg. 510A, Upton, NY 11973, USA\\
        E-mail: \email{nmueller@bnl.gov}}

\author{\speaker{Raju Venugopalan}\\
        Physics Department, Brookhaven National Laboratory, Bldg. 510A, Upton, NY 11973, USA\\
        E-mail: \email{raju@bnl.gov}}

\abstract{In heavy-ion collisions, an interesting question of phenomenological relevance is how the chiral imbalance generated at early times persists through a fluctuating background of sphalerons in addition to other "non-anomalous" interactions with the QGP. To address this question, we construct a relativistic chiral kinetic theory using the world-line formulation of quantum field theory. We outline how Berry's phase arises in this framework, and how its effects can be clearly distinguished from those arising from the chiral anomaly. We further outline how this framework can be matched to classical statistical simulations at early times and to anomalous chiral hydrodynamics at late times.}

\FullConference{Critical Point and Onset of Deconfinement - CPOD2017\\
		7-11 August, 2017\\
		The Wang Center, Stony Brook University, Stony Brook, NY}

\begin{document}

\section{Introduction}
Experimental searches for messengers of CP- and P-odd phenomena in ultra-relativistic heavy ion collisions have aroused much interest and are a prime motivation for significant theoretical effort in this area in recent years. Topological sphaleron transitions~\cite{Klinkhamer:1984di,Dashen:1974ck,Soni:1980ps,Boguta:1983xs,Forgacs:1983yu} were first conjectured to exist in the context of electroweak baryogenesis where they were inferred to be responsible for the matter-antimatter asymmetry in the universe~\cite{Sakharov:1967dj,Riotto:1999yt,Cohen:1993nk,Rubakov:1996vz}. In QCD, these topological 
transitions generate a Chiral Magnetic Effect~\cite{Kharzeev:2007jp,Fukushima:2008xe,Kharzeev:2015znc} in heavy-ion collisions, whereby a combination of CP- and P-odd effects and
strong external (Abelian) magnetic fields can lead to  correlations between charged particles~\cite{Kharzeev:2015znc,Skokov:2016yrj}. However the short lifetime of the external magnetic fields \cite{Skokov:2009qp,Deng:2012pc} suggests that  the CME is largest during the early stages of a collision where the system is far off-equilibrium. The matter at these early times is a strongly correlated "Glasma"-state~\cite{Lappi:2006fp,Gelis:2006dv,Blaizot:2016qgz} formed from the decay of highy occupied Color Glass Condensates~\cite{McLerran:1993ni,McLerran:1993ka,Gelis:2010nm} and its dynamics is captured by ab initio classical-statistical real time lattice simulations ~\cite{Berges:2013eia,Berges:2013fga,Gelis:2016upa}.  These methods can be extended to compute off-equilibrium sphaleron rates in the Glasma~\cite{Mace:2016svc}, which are seen to be significantly larger than the rate of sphaleron transitions in the thermalized Quark-Gluon Plasma \cite{Moore:2010jd}. 

First simulations of dynamical fermions in the gauge fields generating these sphaleron transitions show unambiguously that a CME develops in the presence of the external magnetic field~\cite{Mueller:2016ven,Mace:2016shq,Mace:2017wcl}.
The classical-statistical framework breaks down when occupancies decrease below unity due to the expansion of the overoccupied Glasma.  In this
dilute regime, the results of real-time lattice simulations must be understood as providing the initial conditions for kinetic descriptions
for a weakly coupled dynamical fluid of quasi-particles. Following the space-time evolution of the CME therefore requires that one develops a Boltzmann transport theory for
relativistic chiral fermions which also accounts for its interactions with the topological transitions in the evolving fluid. Such an {\it ab initio} framework is of the utmost importance for systematic analyses of the CME phenomenology, since it can be matched at later times to (anomalous) hydrodynamics ~\cite{Son:2009tf,Nair:2011mk,Gursoy:2014aka,Hongo:2013cqa,Karabali:2014vla,Hirono:2014oda,
Yin:2015fca}.  Several suggestions for a chiral kinetic theory have been made, utilizing models of point particles with
a Berry monopole~\cite{Berry:1984jv}; the latter has been shown to arise in an adiabatic limit for either massive nonrelativistic particles or for massless relativistic particles with a large chemical potential~\cite{Son:2012wh,Stephanov:2012ki,Son:2012zy,Chen:2013iga,Chen:2014cla,Stone:2013sga,Dwivedi:2013dea,Manuel:2014dza,Stone:2014fja,Manuel:2015zpa,Chen:2015gta,Sun:2016nig,Hidaka:2016yjf}.
This framework can be extended to construct an "anomalous B\"odeker theory" \cite{Bodeker:1999ey,Litim:1999id}, which can then be matched to classical-statistical simulations at early times in heavy-ion collisions and to anomalous hydrodynamics at late times. 
It is not clear that early time dynamics in heavy-ion collisions respects these asymptotics and that the inclusion of a Berry term in kinetic descriptions is justified in this context. On the other hand, the physics of the chiral anomaly is ubiquitous and it is essential that chiral kinetic theory frameworks incorporate this physics. An important development is due to \cite{Akamatsu:2013pjd,Akamatsu:2014yza,Yamamoto:2015gzz,Akamatsu:2015kau,Yamamoto:2017uul}, highlighting the possible applicability of such frameworks to astrophysical situations~\cite{Charbonneau:2009ax,Dvornikov:2016gdo,Kaplan:2016drz}. 

In this talk, we will  give an overview of our recent work \cite{Mueller:2017lzw,Mueller:2017arw} in developing a consistent Lorentz covariant transport theory within the world-line 
approach to quantum field theory~\cite{Polyakov:1987ez,Strassler:1992zr,Mondragon:1995va,Mondragon:1995ab,JalilianMarian:1999xt,Schubert:2001he,Bastianelli:2006rx,Hernandez:2008db,Corradini:2015tik}. 
A central ingredient, going back to seminal work by Berezin and Marinov \cite{Berezin:1976eg}
and related contemporaneous work~\cite{Ohnuki:1978jv,Balachandran:1976ya,Balachandran:1977ub,Barducci:1976xq, Galvao:1980cu,Barducci:1982yw,Brink:1976uf, Gitman:1996wk,Gavrilov:2001hk,Naka:1979pn} is the description of spin, color and other internal symmetries via Grassmann variables. The outline of the manuscript is as follow.  In section~\ref{sec:introWL}, we will give a short overview over the world-line representation of the one-loop effective action for fermions. Subsequently,  we
connect the emergence of the chiral anomaly to the existence of fermionic zero modes in the world-line path integral. In section \ref{sec:adiabatic}, we show systematically how Berry's phase arises in our framework. Our explicit computation clearly demonstrates that the topology of the chiral anomaly
and that of Berry's phase  are distinct and should not be confused with each other. For instance, the latter  vanishes when we relax the adiabaticity assumption while the former is robust. In section \ref{sec:CKT}, we outline the elements of a many-body phase space description for relativistic fermions, wherein spin is represented by anti-commuting Grassman coordinates. 

\section{World-line formalism}\label{sec:introWL}
We will briefly here review the world-line representation of quantum field theory
and refer the reader to \cite{Mueller:2017lzw,Mueller:2017arw} for more details.  
We shall restrict ourselves to QED; the extension to QCD
is straightforward. The Euclidean action for massless Dirac fermions in a background vector field $A$ and an auxilliary axial-vector field $B$ is 
\begin{align}
S[A,B]=\int d^4x\;\bar{\psi}\left(i\slashed{\partial}+\slashed{A}+\gamma_5\slashed{B} \right)\psi\,,\label{eq:classicalaction}
\end{align}
The fermion one-loop effective action is given by the fermion determinant $-W[A,B]=\log\det(\theta)$, which can be split 
into a real and and an imaginary part, the latter corresponding to the phase of the determinant. This phase is well known to be responsible for the chiral anomaly \cite{AlvarezGaume:1983ig,AlvarezGaume:1983at}. Remarkably, one can obtain heat-kernel expressions for {\it both} the real and the imaginary part
of the effective action in the world-line formalism~ \cite{D'Hoker:1995ax,D'Hoker:1995bj}, which are expressed through quantum mechanical path integrals of a set of bosonic ($x^\mu$) and anticommuting Grassmann variables $(\psi^\mu,\psi^5,\psi^6)$ quantized on a closed loop.  The real part of the effective action in this formalism can be expressed as 
\begin{align}
W_\mathbb{R}=\frac{1}{8}\int\limits_0^\infty\frac{dT}{T}\mathcal{N}\int\limits_{P}
\mathcal{D}x\int\limits_{AP}\mathcal{D}\psi\;\text{tr}\exp{\Big\{-\int\limits_0^Td\tau\;\mathcal{L}(\tau)\Big\}}\label{eq:realpart}.
\end{align}
where the point particle Lagrangian, $\mathcal{L}=\text{diag}(\mathcal{L}_{L},\mathcal{L}_{R})$,  is given by 
\begin{align}\label{eq:lagrangianexplicit}
&\mathcal{L}_{L/R}=\frac{\dot{x}^2}{2\epsilon}+\frac{1}{2}\psi_a\dot{\psi}_a -i\dot{x}_\mu (A\pm B)_\mu-\frac{i\mathcal{E}}{2}\psi_\mu\psi_\nu F_{\mu\nu}[A\pm B]\,,
\end{align}
with $a=1,\cdots,6$. Here $\epsilon$ is an einbein parameter related to the reparametrization invariance of the world-line. The imaginary part of the effective action can likewise be expressed as a path integral with the same point particle Lagrangian, 
but with a few key differences. Firstly, the imaginary part of the effective action contains a trace insertion
which turns the anti-periodic boundary conditions for the Grassmann variables $\psi$ into periodic ones, thus
being responsible for the existence of fermionic zero modes. Further, this part of the effective action contains an integral over a variable $\alpha$ which has the limits $\pm 1$.  The point particle Lagrangian is identical to that in \Eq{eq:lagrangianexplicit} except that $B\rightarrow \alpha B$. For $\alpha\neq \pm 1$, the imaginary part of the effective action violates chiral symmetry explicitly. The anomaly relation
can be computed directly by varying the imaginary part of the effective action with respect to $B$ and subsequently setting $B=0$: 
\begin{align}\label{eq:finalanomaly}
\partial_\mu \langle j^5_\mu(y)\rangle\equiv
\partial_\mu\frac{i\delta W_\mathbb{I}}{\delta B_\mu(y)}\Big|_{B=0}=-\frac{1}{16\pi^2}\epsilon^{\mu\nu\rho\sigma}F_{\mu\nu}(y)F_{\rho\sigma}(y)\, .
\end{align}
Details of these derivations are worked out in \cite{Mueller:2017lzw,Mueller:2017arw}. Our result clearly shows that the chiral anomaly arises from fermionic zero modes, that uniquely contribute to the imaginary part of the effective action and are absent in the real part of the effective action.

\section{Berry's phase from the real part of the fermion effective action}\label{sec:adiabatic}
In this section, we will investigate the consequences of taking the limit of large masses or large chemical potential in the real part of the effective action and show how Berry's phase further arises in an adiabatic approximation. This cleanly illustrates that the topology of Berry's phase is unrelated to that of the anomaly, as argued previously by Fujikawa and collaborators in more specialized contexts~\cite{Deguchi:2005pc,Fujikawa:2005tv,Fujikawa:2005cn}.

We will first continue the world-line Lagrangian in \Eq{eq:lagrangianexplicit} to Minkowskian metric $g=\text{diag}(-,+,+,+)$. 
To ensure that excitations follow the correct spectrum dictated by the Dirac equation, we introduce the so-called helicity
constraint $\frac{\dot{x}_\mu\psi^\mu}{2\mathcal{E}}+m\psi_5=0$ and a corresponding anti-commuting Lagrange multiplier. Generalizing our discussion to the case of
a massive fermion, and assuming proper time gauge, $\tau=ct\sqrt{1-(\mathbf{v}/c)^2}$, $\sqrt{-\dot{z}^2}=1$, where $\dot{z}^2=(dx^\mu/d\tau)^2$, we find~\cite{Mueller:2017arw} 
\begin{align}
\label{eq:NRLagrangean1}
&\mathcal{L}= -\frac{m_R \,c \,z}{2}\left( 1+\frac{m^2}{m_R^2}\right)+\frac{i}{2}\left(\boldsymbol{\psi} \dot{\boldsymbol{\psi}}- \psi_0\dot{\psi}_0 \right) 
+\frac{\dot{x}_\mu A^\mu(x)}{c} - \frac{iz}{m_R\,c}\,\psi^0 F_{0i}\psi^i- \frac{iz}{2m_R\,c}\,\psi^i F_{ij}\psi^j \, ,
\end{align}
where $m_R^2=m^2+i\psi\cdot F\cdot \psi$. To take the nonrelativistic limit, we expand the action $S=\int d\tau \mathcal{L}=\int dt\,c\sqrt{1-(\mathbf{v}/c)^2} \mathcal{L}$ in powers of
$(v/c)^2$, to obtain 
\begin{align}
\label{eq:NRLagrangean2}
\mathcal{L}_{NR}&=-mc^2+\frac{1}{2}m \boldsymbol{v}^2+\frac{i}{2}\left(\boldsymbol{\psi} \dot{\boldsymbol{\psi}}- \psi_0\dot{\psi}_0 \right) - A^0 +\frac{\boldsymbol{v}}{c}\cdot\boldsymbol{A}
+\frac{ \boldsymbol{S}\cdot(\left[ {\boldsymbol{v}/c}-{\boldsymbol{A}/(mc^2)}\right]\times\boldsymbol{E})}{mc}+\frac{\boldsymbol{S}\cdot\boldsymbol{B}}{m} + O\Bigg(\frac{v^3}{c^3}\Bigg)\,.
\end{align}
The corresponding Hamiltonian is familiar to us from atomic physics: 
\begin{align}\label{eq:NRhamilt}
H\equiv mc^2&+\frac{\left(\boldsymbol{p}-\frac{\boldsymbol{A}}{c}\right)^2}{2m}+A^0(x)
-\frac{ \boldsymbol{S}\cdot(\left[ {\boldsymbol{v}/c}-{\boldsymbol{A}/(mc^2)}\right]\times\boldsymbol{E})}{2mc}-\frac{\boldsymbol{B}\cdot\boldsymbol{S}}{m}\,.
\end{align}
A similar expression is found for the case of a large chemical potential $\mu \gg m$, where the chemical potential 
takes over the role of the mass in \Eq{eq:NRhamilt}. The world-line path integral for the \textit{real} part of the fermion
effective action in the nonrelativistic adiabatic limit can be simplified to read as
\begin{align}\label{eq:Berrypath}
W_\mathbb{R}=\int \mathcal{D}x\mathcal{D}p\;\exp{\Big(i \int dt\;\Big[\dot{\mathbf{x}}\cdot\mathbf{p}-\tilde {H}\Big] \Big)} \,,
\end{align}
with ${\tilde H} = mc^2+\frac{(\boldsymbol{p}-\boldsymbol{A}/c)^2}{2m}+A^0(x)-\dot{\mathbf{p}}\cdot\boldsymbol{\mathcal{A}}(\boldsymbol{p})$. The last term includes the well known Berry phase $\boldsymbol{\mathcal{A}}(\boldsymbol{p})$ $\equiv -i \langle \psi^+(\boldsymbol{p})|\boldsymbol{\nabla}_p| \psi^+(\boldsymbol{p})\rangle$. In deriving this expression, the adiabatic assumption consisted of assuming that the particle's spin was slaved to the direction of the magnetic field and that spin flips are suppressed. Comparing \Eq{eq:Berrypath} to the results of \cite{Son:2012wh,Stephanov:2012ki,Son:2012zy,Chen:2013iga},
we should note a few points: While the emergence of the anomaly was tied to the existence of a Berry phase in \cite{Son:2012wh,Stephanov:2012ki,Son:2012zy,Chen:2013iga}, we see here that we can recover the very same Berry monopole from the
real part of the effective action while the anomaly is tied to the imaginary part.  As stated earlier, our results agree with observations
made by Fujikawa and collaborators in more specific contexts \cite{Deguchi:2005pc,Fujikawa:2005tv,Fujikawa:2005cn,Fujikawa:2017ych}. 
\section{Towards Chiral Kinetic Theory}\label{sec:CKT}
The world-line approach has been employed previously~\cite{JalilianMarian:1999xt} for spinless colored particles to derive the B\"odeker effective kinetic theory for finite temperature QCD~\cite{Bodeker:1999ey,Litim:1999id}. Starting from the Schwinger-Keldysh formulation of the world-line approach,  a quasi-classical many-body description of point particles was obtained by taking the saddle point of the world-line action. The initial density
matrix in the Schwinger-Keldysh (SK) formulation thus represents a stochastic ensemble of classical phase space configurations. For the case of spin, the first step is to identify the 
extended semi-classical phase space to include spinning degrees of freedom; this is analogous to the case of colored particles where the phase space was extended to include their color charge. The natural variables in the spinning case are the Grassmannian variables $\psi$. This gives us  $(x^\mu,P^\mu)\rightarrow(x^\mu,P^\mu,\psi^\mu,\psi_5)$; as discussed previously in \cite{Mueller:2017arw}, $\psi_6$ is not dynamical and is therefore dropped.

The dynamics underlying the microscopic phase space distribution for $N$ particles $f(x,P,\psi)$ 
is governed by the many-body Hamiltonian,
\begin{align}\label{eq:MassiveHamiltonian}
H=\frac{\epsilon}{2}\left(P^2 + m^2 + i \psi^\mu F_{\mu\nu}\psi^\nu \right) + \frac{i}{2}\left(P_\mu \psi^\mu + m\psi_5 \right)\chi
\end{align}
where $P^\mu\equiv p^\mu-A^\mu$ and summation over all particles is implied. In this expression, $\epsilon$ is a commuting Lagrange
multiplier ("einbein") enforcing the mass shell condition while $\chi$ is an anti-commuting Lagrange
multiplier related to the Dirac equation. Both represent gauge parameters with respect to first class constraints and must be fixed before extracting 
physical results. The dynamics of the system can be understood from
the Liouville equation~\cite{Mueller:2017lzw}
\begin{align}\label{eq:Liouville}
0=\{f,H \}=f\Big( \frac{\overleftarrow{\partial}}{\partial x^\mu}\dot{x}^\mu +\frac{\overleftarrow{\partial}}{\partial P^\mu}\dot{P}^\mu +\frac{\overleftarrow{\partial}}{\partial \psi^\mu}\dot{\psi}^\mu +\frac{\overleftarrow{\partial}}{\partial \psi_5}\dot{\psi}_5\Big)\,,
\end{align}
where
\begin{align}
\dot{P}^\mu & =\epsilon P_\alpha F^{\mu\alpha}- \frac{i\epsilon}{2}\psi^\alpha \partial^\mu F_{\alpha\beta}\psi^\beta + \frac{i}{2} F^{\mu\alpha} \psi_\alpha\,\chi\,,\\
\dot{x}^\mu &= \epsilon P^\mu + \frac{i}{2}\psi^\mu \,\chi\,,\\
\dot{\psi}^\mu &= \epsilon F^{\mu\alpha}\psi_\alpha + \frac{P^\mu}{2}\chi\,,\\
\dot{\psi}_5 & = \frac{m}{2}\chi\,.
\end{align}
These equations of motion are generalizations~\cite{Barducci:1976xq}  of the Bargmann-Michel-Telegdi (BMT) equations \cite{Bargmann:1959gz} for spinning particles. For QCD, their equivalents are the Wong equations~\cite{Wong:1970fu}.

Important conceptual issues must be addressed in order to turn this Liouville equation into a transport theory
for fermions whereby both effects of the chiral anomaly and of particle scattering are accounted for. Firstly, since it is impractical to follow every single phase space trajectory via \Eq{eq:Liouville}, a stochastic approach must be
developed that converts the microscopic phase space distribution $f$ into a macroscopic one-particle {\it probability distribution} $\bar{f}$.
Such an analysis should yield the scattering terms in the generalized Boltzmann equation for $\bar{f}$.
The Wigner transform of the initial density matrix in the SK path integral represents a Gibbs ensemble of phase space distributions
in the saddle point limit. This motivates the definition of the distribution functions as
\begin{align}
f\equiv \langle f \rangle + \delta f= \bar{f}+ \delta f
\end{align} 
and similarly for the gauge fields within the SK path integral:
\begin{align}
A^\mu\equiv \langle A^\mu \rangle + \delta A^\mu= \bar{A}^\mu+ \delta A^\mu\,.
\end{align} 
For QED, this also implies $F^{\mu\nu}=\bar{F}^{\mu\nu}+\delta F^{\mu\nu}$. Taking into consideration the fact that the Liouville equation is 
 embedded in the SK path integral, it can be split into separate contributions for the background and fluctuations. The former is given by
\begin{align}
\bar{f}\Big(\frac{\overleftarrow{\partial}}{\partial x^\mu}&\Big[ 
\epsilon P^\mu + \frac{i}{2}\psi^\mu \chi\Big] + \frac{\overleftarrow{\partial}}{\partial P^\mu}\Big[  \epsilon \bar{F}^{\mu\alpha} P_\alpha - \frac{i \epsilon}{2}\psi^\alpha \partial^\mu \bar{F}_{\alpha\beta} \psi^\beta \nonumber\\&+ \frac{i}{2}\bar{F}^{\mu\alpha}\psi_\alpha \chi\Big]
+\frac{\overleftarrow{\partial}}{\partial \psi^\mu}\Big[ 
\epsilon \bar{F}^{\mu\alpha}\psi_\alpha + \frac{P^\mu}{2}\chi
\Big]+\frac{\overleftarrow{\partial}}{\partial \psi_5}\Big[ \frac{m}{2}\chi\Big]\Big)=C[\delta f, \delta F]\,,
\label{Eq:lioudetails2}
\end{align}
where
\begin{align}
C[\delta f, \delta F]\equiv -\epsilon \langle \delta f\frac{\overleftarrow{\partial}}{\partial P^\mu}\,\delta F^{\mu\alpha} \rangle P_\alpha -
\epsilon \langle \delta f\frac{\overleftarrow{\partial}}{\partial \psi^\mu}\,\delta F^{\mu\alpha} \rangle & \psi_\alpha+\frac{i \epsilon}{2} \langle \delta f\frac{\overleftarrow{\partial}}{\partial P^\mu}\,\partial^\mu \delta F_{\alpha\beta}  \rangle \psi^\alpha \psi^\beta \nonumber\\&- \frac{i}{2}\langle \delta f \frac{\overleftarrow{\partial}}{\partial P^\mu} \delta F^{\mu\alpha} \rangle \psi_\alpha \chi\label{eq:scatteringterms}\,.
\end{align}
The equation for the fluctuations $\delta f$ is in general a complicated expression involving products of $\delta f$ and $\delta F^{\mu\nu}$
to higher orders. It serves as a generating functional for equations for higher moments of fluctuations, resulting in an infinite hierarchy 
of coupled equations, analogous to the BBGKY hierachy \cite{BBGKY,deGroot,Landau}. As noted previously, since  classical-statistical simulations are sufficient to describe single particle distributions for $\bar{f}\geq 1$, we can simplify the above to consider the dilute limit of $\delta f\ll \bar{f} \lessapprox 1$, where the collision term for the equation for fluctuations  is sub-dominant. Therefore keeping only linear terms in both  $\delta f$ and $\delta F^{\mu\nu}$, we obtain the evolution equation for $\delta f$ to be 
\begin{align}
\delta f\Big(\frac{\overleftarrow{\partial}}{\partial x^\mu}
\Big[\epsilon P^\mu +& \frac{i}{2}\psi^\mu \chi\Big]
\frac{\overleftarrow{\partial}}{\partial P^\mu}\Big[ \epsilon \bar{F}^{\mu\alpha}P_\alpha -\frac{i\epsilon}{2}\psi^\alpha \partial^\mu \bar{F}_{\alpha\beta}\psi^\beta \nonumber\\&+\frac{i}{2}\bar{F}^{\mu\alpha}\psi_\alpha \chi\Big]+\frac{\overleftarrow{\partial}}{\partial \psi^\mu}\Big[ \epsilon \bar{F}^{\mu\alpha}\psi_\alpha + \frac{P^\mu}{2}\chi\Big]+\frac{\overleftarrow{\partial}}{\partial \psi_5}\Big[ \frac{m}{2}\chi\Big] 
\Big)=K[\delta F]\,,
\label{Eq:lioudetails3} 
\end{align}  
where
\begin{align}
K[\delta F]\equiv -\bar{f}\Big(\frac{\overleftarrow{\partial}}{\partial P^\mu}\Big[ \epsilon \delta{F}^{\mu\alpha}P_\alpha -\frac{i\epsilon}{2}\psi^\alpha \partial^\mu \delta{F}_{\alpha\beta}\psi^\beta +\frac{i}{2}\delta{F}^{\mu\alpha}\psi_\alpha \chi\Big]
+\frac{\overleftarrow{\partial}}{\partial \psi^\mu}\Big[\epsilon \delta F^{\mu\alpha}\psi_\alpha\Big] \Big) \,.
\end{align}
The coupled set of equations \Eq{Eq:lioudetails2} and \Eq{Eq:lioudetails3} constitute the key ingredients of the chiral kinetic theory that we wish to construct. Much work remains to flesh out the structure of these equations and write them down in a fashion that is of practical use. This work is in progress. 

Missing from our preceding kinetic theory discussion is the contribution from the chiral anomaly, which was the principal motivation for this work. However we know how to include the anomaly in the SK framework since its contribution is contained in the fermion zero modes in the SK path integral. Finally, since the Liouville equation must be evaluated inside the world-line path integral, it is beneficial to split the distribution function $f$ into distributions for positive and negative chiralities. 
In the case of chiral fermions, the Hamiltonian in \Eq{eq:MassiveHamiltonian} is modified to include the Weyl constraints
\begin{align}
\frac{1}{2}(\gamma\cdot p)(1\pm \gamma^5)\Psi=0\,.\label{eq:WeylConstraint}
\end{align}
The conservation and non-conservation of chiral distribution functions can  then be understood
easily when the above discussion is generalized to the case of chiral Weyl fermions. Work in this direction is promising and will be reported on separately. 

\section{Acknowledgments}
The authors are supported by the U.S. Department of Energy,
Office of Science, Office of Nuclear Physics, under contract No. DE- SC0012704, and within the framework of the Beam Energy Scan Theory (BEST) Topical Collaboration.

\end{document}